\documentclass[conference]{IEEEtran}
\IEEEoverridecommandlockouts
\usepackage{acro}[=v2]

\NewDocumentCommand{\acro}{m o m o}
{%
	\IfValueTF{#2}{%
		\IfValueTF{#4}{%
			\DeclareAcronym{#1}{short={#2},long={#3},#4}
		}{%
			\DeclareAcronym{#1}{short={#2},long={#3}}
		}
	}{%
		\IfValueTF{#4}{%
			\DeclareAcronym{#1}{short={#1},long={#3},#4}
		}{%
			\DeclareAcronym{#1}{short={#1},long={#3}}
		}
	}
}

\acro{2D}{two-dimensional}
\acro{3D}{three-dimensional}
\acro{4D}{four-dimensional}
\acro{6D}{six-dimensional}
\acro{1G}{first generation}
\acro{2G}{second generation}
\acro{3G}{third generation}
\acro{4G}{fourth generation}
\acro{5G}{fifth generation}
\acro{5GC}{5G core network}
\acro{5G-StoRM}{5G stochastic radio channel for dual mobility}
\acro{6G}{sixth generation}
\acro{3GPP}{3rd generation partnership project}
\acro{3GPP2}{3rd generation partnership project 2}
\acro{A2A}{air-to-air}
\acro{A2G}{air-to-ground}
\acro{AA}{antenna array}
\acro{AC}{admission control}
\acro{AD}{attack-decay}
\acro{AE}{antenna element}
\acro{AF}{amplify and forward}
\acro{ABS}{almost blank subframe}
\acro{ACF}{autocorrelation function}
\acro{ADSL}{asymmetric digital subscriber line}
\acro{AHW}{alternate hop-and-wait}
\acro{AI}{artificial intelligence}
\acro{AMC}{adaptive modulation and coding}
\acro{AoA}{angle of arrival}
\acro{AoD}{angle of departure}
\acro{AP}{access point}
\acro{APA}{adaptive power allocation}
\acro{API}{application protocol interface}
\acro{ARQ}{automatic repeat request}
\acro{AT}{averaged time}
\acro{ARMA}{autoregressive moving average}
\acro{ASE}{average squared error}
\acro{ASC}{adaptive satisfaction control}
\acro{ATES}{adaptive throughput-based efficiency-satisfaction trade-off}
\acro{AWGN}{additive white gaussian noise}
\acro{B5G}{beyond 5G}
\acro{BAP}{backhaul adaptation protocol}
\acro{BB}{branch and bound}
\acro{BC}{branch and cut}
\acro{BD}{block diagonalization}
\acro{BER}{bit error rate}
\acro{BF}{best fit}
\acro{BL}{bit loading}
\acro{BLER}{block error rate}
\acro{BLPC-1}{bit loading with power constraint in hop 1}
\acro{BLPC-2}{bit loading with power constraint in hop 2}
\acro{BPC}{binary power control}
\acro{BPSK}{binary phase-shift keying}
\acro{BRA}{balanced random allocation}
\acro{BS}{base station}
\acro{BSP}{\acs*{BS} placement}
\acro{BSR}{buffer status report}
\acro{BoI}{band of interest}
\acro{C-link}{control link}
\acro{CAP}{combinatorial allocation problem}
\acro{CAPEX}{capital expenditure}
\acro{CB}{contextual bandit}
\acro{CBF}{coordinated beamforming}
\acro{CBR}{constant bit rate}
\acro{CBS}{class based scheduling}
\acro{CC}{congestion control}
\acro{CCL}{common cell list}
\acro{CDF}{cumulative distribution function}
\acro{CDL}{clustered delay line}
\acro{CDMA}{code-division multiple access}
\acro{CIR}{channel impulse response}
\acro{CH}{channel hardening}
\acro{CHO}{conditional handover}
\acro{C-RAN}{cloud-based radio access network}
\acro{CL}{closed loop}
\acro{CLT}{central limit theorem}
\acro{CLPC}{closed loop power control} 
\acro{CN}{core network}       
\acro{CNR}{channel-to-noise ratio}
\acro{CP}{control plane}
\acro{CPA}{cellular protection algorithm}
\acro{CPICH}{common pilot channel}
\acro{CoMP}{coordinated multi-point}
\acro{CQI}{channel quality indicator}
\acro{CRE}{cell range expansion}
\acro{CRM}{constrained rate maximization}
\acro{CRN}{cognitive radio network}
\acro{C-RNTI}{cell radio network temporary identifier}
\acro{CRRM}{centralized/common radio resource management}
\acro{CRS}{cell-specific reference signal}
\acro{CS}{coordinated scheduling}
\acro{CSI}{channel state information}
\acro{CSI-RS}{channel state information reference signal}
\acro{CTS}{clear to send}
\acro{CU}{centralized unit}
\acro{CUE}{cellular user equipment}
\acro{CWND}{congestion window size}
\acro{D2D}{device-to-device}
\acro{D2N}{drone-to-network}
\acro{DAA}{drone antenna array}
\acro{DAG}{directed acyclic graph}
\acro{DBS}{drone mounted base station}
\acro{DC}{dual connectivity}
\acro{DCA}{dynamic channel allocation}
\acro{DE}{differential evolution}
\acro{DF}{decode and forward}
\acro{DFT}{discrete fourier transform}
\acro{DIST}{distance}
\acro{DL}{downlink}
\acro{DMA}{double moving average}
\acro{DMRS}{demodulation reference signal}
\acro{D2DM}{\acs*{D2D} mode}
\acro{DMS}{\acs*{D2D} mode selection}
\acro{DP}{data plane}
\acro{DPC}{dirty paper coding}
\acro{DRA}{dynamic resource assignment}
\acro{DSA}{dynamic spectrum access}
\acro{DSM}{delay-based satisfaction maximization}
\acro{DU}{distributed unit}
\acro{E2E}{end-to-end}
\acro{ECC}{electronic communications committee}
\acro{EDF}{earliest deadline first}
\acro{EE}{energy efficiency}
\acro{EFLC}{error feedback based load control}
\acro{EI}{efficiency indicator}
\acro{e-ICIC}{enhanced inter-cell interference coordination}
\acro{eMBB}{enhanced mobile broadband}
\acro{ETSI}{European Telecommunications Standards Institute}
\acro{EM}{electromagnetic}
\acro{eNB}{evolved node B}
\acro{EXP}{exponential}
\acro{EPA}{equal power allocation}
\acro{EPC}{evolved packet core}
\acro{EPS}{evolved packet system}
\acro{E-UTRAN}{evolved universal terrestrial radio access network}
\acro{ES}{exhaustive search}
\acro{FCP}{fundamental counting principle}
\acro{FCA}{flow control algorithm}
\acro{FD}{full duplex}
\acro{FDD}{frequency division duplex}
\acro{FDM}{frequency division multiplexing}
\acro{FDMA}{frequency division multiple access}
\acro{FER}{frame erasure rate}
\acro{FIFO}{first in first out}
\acro{FoV}{field-of-view}
\acro{FF}{fast fading}
\acro{FL}{federated learning}
\acro{FR}{frequency range}
\acro{FRS}[FS]{fast-rat scheduling}
\acro{FS}{fast switching}
\acro{FSB}{fixed switched beamforming}
\acro{FST}{fixed \acs*{SNR} target}
\acro{FT}{fourier transform}
\acro{FTP}{file transfer protocol}
\acro{Fwd}{forwarding}
\acro{GA}{genetic algorithm}
\acro{GAP}{generalized assignment problem}
\acro{GAP-MQ}{generalized assignment problem with minimum quantities}
\acro{GATES}{generalized adaptive throughput-based efficiency-satisfaction trade-off}
\acro{GBR}{guaranteed bit rate}
\acro{GLR}{gain to leakage ratio}
\acro{gNB}{gNodeB}
\acro{GOS}{generated orthogonal sequence}
\acro{GP}{gaussian process}
\acro{GPL}{GNU general public license}
\acro{GPS}{global positioning system}
\acro{GRP}{grouping}
\acro{GR}{group report}
\acro{GSM}{global system for mobile communications}
\acro{GTEL}{wireless telecommunications research group}
\acro{HAP}{high altitude platform}
\acro{HARQ}{hybrid automatic repeat request}
\acro{HBF}{hybrid beamforming}
\acro{HCPP}{hardcore point process}
\acro{HetNet}{heterogeneous network}
\acro{HD}{half duplex}
\acro{HF}{high-frequency}
\acro{HH}{hughes-hartogs}
\acro{HardH}[HH]{hard handover}
\acro{HMS}{harmonic mode selection}
\acro{HO}{handover}
\acro{HOL}{head of line}
\acro{HPBW}{half power beamwidth}
\acro{HSDPA}{high speed downlink packet access}
\acro{HSR}{high-speed railway}
\acro{HSPA}{high speed packet access}
\acro{HTTP}{hypertext transfer protocol}
\acro{IAB}{integrated access and backhaul}
\acro{ICMP}{internet control message protocol} 
\acro{ICI}{intercell interference}
\acro{ICIC}{inter-cell interference coordination}
\acro{ID}{identification}
\acro{i.i.d.}{independent and identically distributed}
\acro{IETF}{internet engineering task force}
\acro{IPC}{individual power constraint}
\acro{ISAC}{integrated sensing and communication}
\acro{UID}{unique identification}
\acro{IID}{independent and identically distributed}
\acro{IIR}{infinite impulse response}
\acro{ILP}{integer linear problem}
\acro{IMT}{international mobile telecommunications}
\acro{INV}{inverted norm-based grouping} 
\acro{IoT}{internet of things}
\acro{IP}{internet protocol}
\acro{IPv6}{internet protocol version 6}
\acro{IRA}{integrated resource allocation}
\acro{IRS}{intelligent reflecting surface}
\acro{ISD}{inter-site distance}
\acro{ISI}{inter symbol interference}
\acro{ISM}{industrial, scientific and medical}
\acro{ITU}{international telecommunication union}
\acro{JOAS}{joint opportunistic assignment and scheduling}
\acro{JOS}{joint opportunistic scheduling}
\acro{JP}{joint processing}
\acro{JRAPAP}{joint rb assignment and power allocation problem}
\acro{JS}{jump-stay}
\acro{JSM}{joint satisfaction maximization}
\acro{KKT}{karush-kuhn-tucker}
\acro{KPI}{key performance indicator}
\acro{LAC}{link admission control}
\acro{LA}{link adaptation}
\acro{LAP}{low altitude platform}
\acro{LBS}{location based service}
\acro{LC}{load control}
\acro{LEB}{load-efficiency-balance}
\acro{LOS}{line of sight}
\acro{LP}{linear programming}
\acro{LSTM}{long short-term memory}
\acro{LTE}{long term evolution}
\acro{LTE-A}{\acs*{LTE}-advanced}
\acro{LTE-Advanced}{\ac{LTE-A}}
\acro{LTE-R}{\ac{LTE} railway}
\acro{LSP}{large scale parameter}
\acro{MADRDPG}{multi-agent deep recurrent deterministic policy gradient}
\acro{MeNB}{master \acs*{ENB}}
\acro{M2M}{machine-to-machine}
\acro{MAC}{medium access control}
\acro{MANET}{mobile ad hoc network}
\acro{MEDS}{method of exact doppler spread}
\acro{MC}{modular clock}
\acro{MCP}{minimal cost power}
\acro{MCS}{modulation and coding scheme}
\acro{MDB}{measured delay based}
\acro{MDI}{minimum \acs*{D2D} interference}
\acro{MDSM}{modified delay-based satisfaction maximization}
\acro{MDU}{max-delay-utility}
\acro{METIS}{mobile and wireless communications enablers for the twenty-twenty information society \acs*{5G}}
\acro{MF}{matched filter}
\acro{MFAP}{mobile femtocell access point}
\acro{MG}{maximum gain}
\acro{MH}{multi-hop}
\acro{mIAB}{mobile \ac{IAB}}
\acro{MILP}{mixed integer linear programming}
\acro{MIMO}{multiple input multiple output}
\acro{MINLP}{mixed integer nonlinear programming}
\acro{MIP}{mixed integer programming}
\acro{MISO}{multiple input single output}
\acro{MIT}{mobility interruption time}
\acro{ML}{machine learning}
\acro{MLWDF}{modified largest weighted delay first}
\acro{MME}{mobility management entity}
\acro{MMF}{max-min fairness}
\acro{mmMAGIC}{millimetre-wave based mobile radio access network for fifth generation integrated communications}
\acro{MMRP}{maximizing the minimal residual power}
\acro{MMRP-LB}{maximizing the minimal residual power with lower bound}
\acro{MMSE}{minimum mean square error}
\acro{mMTC}{massive machine-type communications}
\acro{mmWave}{millimeter wave}
\acro{MN}{master node}
\acro{MOS}{mean opinion score}
\acro{MPF}{multicarrier proportional fair}
\acro{MPRP}{maximization of the product of the residual powers}
\acro{MRA}{maximum rate allocation}
\acro{MR}{maximum rate}
\acro{MRC}{maximum ratio combining}
\acro{MRN}{mobile relay node}
\acro{MRT}{maximum ratio transmission}
\acro{MRUS}{maximum rate with user satisfaction}
\acro{MS}{mode selection}
\acro{MSE}{mean squared error}
\acro{MCG}{master cell group} 
\acro{MSI}{multi-stream interference}
\acro{MT}{mobile termination}
\acro{MTC}{machine-type communication}
\acro{IMS}{\acs*{IP} multimedia subsystem}
\acro{MTSI}{multimedia telephony services over \acs*{IMS}}
\acro{MTSM}{modified throughput-based satisfaction maximization}
\acro{MU-MIMO}{multi-user multiple input multiple output}
\acro{MU}{multi-user}
\acro{Multi-CUT}{multi-cell and multi-user and multi-tier}
\acro{NAS}{non-access stratum}
\acro{NB}{node B}
\acro{nBS}{neighbor base station}
\acro{NCL}{neighbor cell list}
\acro{NCR}{network-controlled repeater}
\acro{NG}{next generation}
\acro{NGC}{next generation core network}
\acro{NLP}{nonlinear programming}
\acro{NLOS}{non-line of sight}
\acro{NMSE}{normalized mean square error}
\acro{NN}{neural network}
\acro{NOMA}{non-orthogonal multiple access}
\acro{NORM}{normalized projection-based grouping}
\acro{NP}{non-polynomial time}
\acro{NR}{new radio}
\acro{NRT}{non-real time}
\acro{NSA}{non-stand-alone}
\acro{NSPS}{national security and public safety services}
\acro{OAM}{operation and maintenance}
\acro{OBF}{opportunistic beamforming}
\acro{OFDMA}{orthogonal frequency division multiple access}
\acro{OFDM}{orthogonal frequency division multiplexing}
\acro{OFPC}{open loop with fractional path loss compensation}
\acro{O2I}{outdoor-to-indoor}
\acro{OL}{open loop}
\acro{OLPC}{open-loop power control}
\acro{OL-PC}{open-loop power control}
\acro{OM}[O\&M]{operational \& maintenance}
\acro{OMA}{orthogonal multiple access}
\acro{OPEX}{operational expenditure}
\acro{ORA}{orthogonal resource allocation}
\acro{ORB}{orthogonal random beamforming}
\acro{OSM}{open street map}
\acro{JO-PF}{joint opportunistic proportional fair}
\acro{OSI}{open systems interconnection}
\acro{PA}{power allocation}
\acro{PAIR}{\acs*{D2D} pair gain-based grouping}
\acro{PAPR}{peak-to-average power ratio}
\acro{P2P}{peer-to-peer}
\acro{PBCH}{physical broadcast channel}
\acro{PBS}{pico base station}        
\acro{PC}{power control}
\acro{PCI}{physical cell \acs*{ID}}
\acro{PDCP}{packet data convergence protocol}
\acro{PDF}{probability density function}
\acro{PDU}{protocol data unit}
\acro{PER}{packet error rate}
\acro{PF}{proportional fair}
\acro{PGRA}{probabilistic graph based resource allocation}
\acro{P-GW}{packet data network gateway}
\acro{PHY}{physical}
\acro{PIN}{positive-intrinsic-negative}
\acro{PL}{pathloss}
\acro{PLR}{packet loss ratio}
\acro{PPP}{Poisson point process}
\acro{PRABE}{power and resource allocation based on quality of experience}
\acro{PRB}{physical resource block}
\acro{PROJ}{projection-based grouping}
\acro{PSD}{power spectral density}
\acro{PSDe}{positive semi-definite}
\acro{ProSe}{proximity services}
\acro{PS}{packet scheduling}
\acro{PSO}{particle swarm optimization}
\acro{PSS}{primary synchronization signal}
\acro{PTAS}{polynomial-time approximation scheme}
\acro{PTP}{point-to-point}
\acro{PZF}{projected zero-forcing}
\acro{QAM}{quadrature amplitude modulation}
\acro{QHMLWDF}{queue-HOL-MLWDF}
\acro{QoE}{quality of experience}
\acro{QoS}{quality of service}
\acro{QPSK}{quadri-phase shift keying}
\acro{QSM}{queue-based satisfaction maximization}
\acro{QuaDRiGa}{quasi deterministic radio channel generator}
\acro{RACH}{random access}
\acro{RAISES}{reallocation-based assignment for improved spectral efficiency and satisfaction}
\acro{RAN}{radio access network}
\acro{RA}{resource allocation}
\acro{RAP}{rb assignment problem}
\acro{RAT}{radio access technology}[long-plural-form={radio access technologies}]
\acro{RATE}{rate-based}
\acro{RAU}{remote antenna unit}
\acro{RB}{resource block}
\acro{RBG}{resource block group}
\acro{REF}{reference grouping}
\acro{RET}{remote electrical tilt}
\acro{RF}{radio frequency}
\acro{RL}{reinforcement learning}
\acro{RLC}{radio link control}
\acro{RLF}{radio link failure}
\acro{RM}{rate maximization}
\acro{RMa}{rural macro}
\acro{RMJ-SNR}{region of the minimum joint snr}
\acro{RMEC}{rate maximization under experience constraints}
\acro{RMSE}{root-mean-square error}
\acro{RNC}{radio network controller}
\acro{RND}{random grouping}
\acro{RNN}{recurrent neural network}
\acro{RRA}{radio resource allocation}
\acro{RRM}{radio resource management}
\acro{RSCP}{received signal code power}
\acro{RSRP}{reference signal received power}
\acro{RSRQ}{reference signal received quality}
\acro{RR}{round robin}
\acro{RIS}{reconfigurable intelligent surface}
\acro{RRC}{radio resource control}
\acro{RSSI}{received signal strength indicator}
\acro{RT}{real time}
\acro{RTR}[RT]{ray-tracing}
\acro{RTS}{request to send}
\acro{RU}{resource unit}
\acro{RUNE}{rudimentary network emulator}
\acro{RV}{random variable}
\acro{Rx}{receiver}
\acro{RZF}{regularized zero-forcing}
\acro{SA}{subcarrier assignment}
\acro{SAC}{session admission control}
\acro{sBS}{serving base station}
\acro{SC}{small cell}
\acro{SCI}{side control information}
\acro{SCon}[SC]{single connectivity}
\acro{SCG}{secondary cell group}
\acro{SCM}{spatial channel model}
\acro{SCS}{subcarrier spacing}
\acro{SC-FDMA}{single carrier - frequency division multiple access}
\acro{SCRV}{spatially consistent random variable}
\acro{SD}{soft dropping}
\acro{SDAP}{service	data adaptation protocol}
\acro{S-D}{source-destination}
\acro{SDPC}{soft dropping power control}
\acro{SDM}{space-division multiplexing}
\acro{SDMA}{space-division multiple access}
\acro{SE}{spectral efficience}
\acro{SF}{shadow fading}
\acro{SMDP}{semi-markov	decision problem}
\acro{SeNB}{secondary \acs*{ENB}}
\acro{SER}{symbol error rate}
\acro{SES}{simple exponential smoothing}
\acro{S-GW}{serving gateway}
\acro{SIC}{successive interference cancellation}
\acro{SelfIC}[SIC]{self interference cancellation}
\acro{SINR}{signal to interference-plus-noise ratio}
\acro{SLNR}{signal to leakage-plus-noise ratio}
\acro{SNN}{strictly non-negative}
\acro{SI}{satisfaction indicator}
\acro{SelfI}[SI]{self interference}
\acro{SIP}{session initiation protocol}
\acro{SISO}{single input single output}
\acro{SIMO}{single input multiple output}
\acro{SIR}{signal to interference ratio}
\acro{SM}{subcarrier matching}
\acro{SMA}{simple moving average}
\acro{SMSE}{spatial-mean-square error}
\acro{SN}{secondary node}
\acro{SNR}{signal to noise ratio}
\acro{SON}{self organizing networks}
\acro{SoA}{state-of-the-art}
\acro{SoS}{sum-of-sinusoids}
\acro{SORA}{satisfaction oriented resource allocation}
\acro{SORA-NRT}{satisfaction-oriented resource allocation for non-real time services}
\acro{SORA-RT}{satisfaction-oriented resource allocation for real time services}
\acro{SP}{signal processing}
\acro{SPF}{single-carrier proportional fair}
\acro{SR}{smart repeater}
\acro{ASR}{advanced \ac{SR}}
\acro{SRA}{sequential removal algorithm}
\acro{SRB1}{signaling radio bearer~1}
\acro{SRS}{sounding reference signal}
\acro{SSB}{synchronization signal block}
\acro{SSP}{small scale parameter}
\acro{SSS}{secondary synchronization signal}
\acro{ST}{spanning tree}
\acro{STTD}{space time transmit diversity}[long-plural-form={space time transmit diversities}]
\acro{SU-MIMO}{single-user multiple input multiple output}
\acro{SU}{single-user}
\acro{tBS}{target base station}
\acro{SVD}{singular value decomposition}
\acro{TCP}{transmission control protocol}
\acro{TDD}{time division duplex}
\acro{TDM}{time division multiplexing}
\acro{TDMA}{time division multiple access}
\acro{TDMed}{time division multiplexed}
\acro{TETRA}{terrestrial trunked radio}
\acro{TP}{transmit power}
\acro{TPC}{total power constraint}
\acro{TR}{technical report}
\acro{TS}{technical specification}
\acro{TTI}{transmission time interval}
\acro{TTR}{time-to-rendezvous}
\acro{TTT}{time-to-trigger}
\acro{TSM}{throughput-based satisfaction maximization}
\acro{TU}{typical urban}
\acro{TV}{television}
\acro{TVWS}{\acs*{TV} white space}
\acro{Tx}{transmitter}
\acro{UAV}{unmanned aerial vehicle}
\acro{UABSC}{user-assisted bearer split control}
\acro{UDP}{user datagram protocol}
\acro{UDN}{ultra-dense networks}
\acro{UE}{user equipment}
\acro{UBA}{\ac{UE}-\ac{BS} association}
\acro{UEPS}{urgency and efficiency-based packet scheduling}
\acro{UFC}{federal university of cear\'{a}}
\acro{ULA}{uniform linear array}
\acro{UL}{uplink}
\acro{UMa}{urban macro}
\acro{UMi}{urban micro}
\acro{UMTS}{universal mobile telecommunications system}
\acro{UP}{user plane}
\acro{UPA}{uniform planar array}
\acro{UPN}{user provided network}
\acro{URA}{uniform rectangular array}
\acro{URI}{uniform resource identifier}
\acro{URLLC}{ultra-reliable low-latency communications}
\acro{URM}{unconstrained rate maximization}
\acro{VBR}{variable bit rate}
\acro{VET}{variable electrical tilt}
\acro{VMR}{vehicle-mounted relay}
\acro{VR}{virtual resource}
\acro{VoIP}{voice over \acs*{IP}}
\acro{WCDMA}{wideband code division multiple access}
\acro{WF}{water-filling}
\acro{WID}{wireless infrastructure drone}
\acro{Wi-Fi}{wireless fidelity}
\acro{WiMAX}{worldwide interoperability for microwave access}
\acro{WINNER}{wireless world initiative new radio}
\acro{WLAN}{wireless local area network}
\acro{WMPF}{weighted multicarrier proportional fair}
\acro{WP}{work package}
\acro{WPF}{weighted proportional fair}
\acro{WSN}{wireless sensor network}
\acro{WWW}{world wide web}
\acro{WFQ}{weighted fair queuing}
\acro{XIXO}{(single or multiple) input (single or multiple) output}
\acro{XPR}{cross-polarization power ratio}
\acro{ZF}{zero-forcing}
\acro{ZMCSCG}{zero mean circularly symmetric complex gaussian}

\usepackage{amsmath,amssymb,amsfonts}
\DeclareMathAlphabet{\mathppl}{T1}{ppl}{m}{it}
\DeclareMathAlphabet{\mathphv}{T1}{phv}{m}{it}
\DeclareMathAlphabet{\mathpzc}{T1}{pzc}{m}{it}

\newcommand{\Set}[1]{\mathcal{\uppercase{#1}}}

\newcommand{\stR}{\Set{R}}

\newcommand{\stU}{\Set{U}}

\usepackage[%
style=ieee,
backend=biber, 
]{biblatex}
\usepackage{siunitx}
\newcommand{\SecRef}[2][]{Section#1~\ref{#2}}
\newcommand{\FigRef}[2][]{Fig.#1~\ref{#2}}

\newcommand{\TabRef}[2][]{Table#1~\ref{#2}}

\usepackage{algorithmic}
\usepackage{subfig}
\usepackage{graphicx}
\usepackage{textcomp}

\def\BibTeX{{\rm B\kern-.05em{\sc i\kern-.025em b}\kern-.08em
    T\kern-.1667em\lower.7ex\hbox{E}\kern-.125emX}}

\usepackage{multirow}
\usepackage{booktabs}
\usepackage{adjustbox}
\usepackage{array}
\usepackage{tabularx}
\newcolumntype{C}{>{\centering\arraybackslash}X}
\newcolumntype{R}{>{\raggedleft\arraybackslash}X}

\usepackage[dvipsnames,svgnames,x11names]{xcolor}
\usepackage{standalone}
\usepackage{tikz}
\usetikzlibrary{arrows.meta} 
\usetikzlibrary{arrows}
\usetikzlibrary{backgrounds}
\usetikzlibrary{calc}
\usetikzlibrary{chains}
\usetikzlibrary{decorations.pathreplacing}
\usetikzlibrary{decorations.text}
\usetikzlibrary{fit}
\usetikzlibrary{intersections}
\usetikzlibrary{matrix}
\usetikzlibrary{positioning}
\usetikzlibrary{scopes}
\usetikzlibrary{shadows}
\usetikzlibrary{shapes.geometric}
\usetikzlibrary{shapes.misc}
\usetikzlibrary{shapes.multipart}
\usetikzlibrary{shapes.symbols}
\usetikzlibrary{shapes}

\graphicspath{%
	{./figs/}
}

\usepackage{pgfplots}
\pgfplotsset{%
	width=0.95*\columnwidth,
	height=0.25\columnwidth, 
	compat=1.14,
	compat/show suggested version=false,
	filter discard warning=false,
	tick label style={font=\footnotesize},
	label style={font=\footnotesize},
	every axis label={font=\footnotesize},
	grid=major,
	grid style={dashed,gray!30},
	cycle list shift=0,
	enlargelimits=false,
	legend style={%
		font=\footnotesize,
		legend cell align=left,
		nodes={inner xsep=2pt,inner ysep=1pt,text depth=0.15em},
	},
}
\usepackage{pgfplotstable}

\newtoggle{tikzexternalize}
\togglefalse{tikzexternalize}

\iftoggle{tikzexternalize}{
	\immediate\write18{mkdir -p ./tikz-external-figs/}
	\usetikzlibrary{external}
	\tikzexternalize[prefix=./tikz-external-figs/,mode=list and make]
	\tikzset{external/system call={pdflatex \tikzexternalcheckshellescape -halt-on-error -interaction=batchmode -jobname "\image" "\texsource"}}
}{
}

\usepackage[final]{changes} 
\usepackage{placeins}
\def\plotwidth{0.99\columnwidth}
\def\plotheight{0.5\columnwidth}

\pgfplotsset{common line style/.style={line width=1pt}}

\pgfplotsset{every axis plot post/.append style={
    every mark/.append style={scale=1.5}
}}

\pgfplotsset{common plots axis options/.style={
	every axis/.append style={
	  legend style={fill=gray!5, fill opacity=0.85, text opacity=1}
	},
	width=\plotwidth,
	height=\plotheight,
	grid=both,
	filter discard warning=false,
	tick label style={font=\footnotesize},
	label style={font=\footnotesize},
	every axis label={font=\footnotesize},
	grid=major,
	grid style={dashed,gray!30},
	cycle list shift=0,
	enlargelimits={true,abs value=1pt},
	ylabel shift = -0.5ex,
	legend style={%
		font=\scriptsize,
		legend cell align=left,
		nodes={inner xsep=3pt,inner ysep=2pt,text depth=0.15em},
	},
	}
}

\pgfplotsset{bar axis options/.style={
		common plots axis options,
		ybar=1pt,
		bar width = 3pt,
		enlarge x limits={true,abs value=5pt},
}}

\pgfplotsset{mcs axis options/.style={
		bar axis options,
		width=0.9\columnwidth,
		height=0.5\columnwidth,
		ymin = 0, ymax = 100,
		xtick={0, 1, ..., 15},
		xticklabels={Total, MCS 1, MCS 2, MCS 3, MCS 4, MCS 5, MCS 6, MCS 7, MCS 8, MCS 9, MCS 10, MCS 11, MCS 12, MCS 13, MCS 14, MCS 15},
		x tick label style={
			font=\tiny,
			xshift = 1ex,
			rotate=45,
			anchor=east,
		},
}}

\pgfplotsset{common marker style/.style={
		mark repeat = 10,
		mark size = 1pt,
		mark options={solid},
}}

\pgfplotsset{scenario0 style/.style={
	common line style,
	common marker style,
	MediumSeaGreen,
	mark=triangle*,
}}

\pgfplotsset{scenario1 style/.style={
	common line style,	
	common marker style,
	SandyBrown,
	mark=square*,
}}

\pgfplotsset{scenario12 style/.style={
    common line style,
    common marker style,
    DodgerBlue,
    mark=*,
}}

\pgfplotsset{scenario2 style/.style={
		common line style,
		common marker style,
		FireBrick,
		mark=x,
		mark size = 2pt,
}}

\pgfplotsset{scenario21 style/.style={
		common line style,
		common marker style,
		DarkSlateGray,
		mark=diamond*,
}}

\pgfplotsset{load bar style/.style={
		DodgerBlue, fill
}}

\pgfplotsset{ack bar style/.style={
		DodgerBlue, fill
}}

\pgfplotsset{nack bar style/.style={
		Red, fill
}}

\def\plotsDataPath{figs/plots/data}

\begin{document}

\title{Impact of Network Deployment on the \\ Performance of NCR-assisted Networks \\

\thanks{This work was supported by Ericsson Research, Sweden, and Ericsson Innovation Center, Brazil, under UFC.51 Technical Cooperation Contract Ericsson/UFC. It was also financed in part by the Coordenação de Aperfeiçoamento de Pessoal de Nível Superior – Brasil (CAPES) – Finance Code 001. The authors acknowledge the partial support of FUNCAP/Universal under grant no. UNI-0210-00043.01.00/23. The work of Victor F. Monteiro was supported by CNPq under Grant 308267/2022-2. The work of Tarcisio F. Maciel was supported by CNPq under Grant 312471/2021-1. The work of Francisco R. M. Lima was supported by FUNCAP (edital BPI) under Grant BP5-0197-00194.01.00/22.}
}

\author{\IEEEauthorblockN{Gabriel C. M. da Silva}
\IEEEauthorblockA{\textit{Federal University of Ceará}\\
Fortaleza, Brazil\\
gabriel@gtel.ufc.br}
\and
\IEEEauthorblockN{Diego A. Sousa}
\IEEEauthorblockA{\textit{Federal Institute of Ceará}\\
Paracuru, Brazil  \\
diego@gtel.ufc.br}
\and
\IEEEauthorblockN{Victor F. Monteiro}
\IEEEauthorblockA{\textit{Federal University of Ceará}\\
Fortaleza, Brazil \\
victor@gtel.ufc.br}
\and
\IEEEauthorblockN{Darlan C. Moreira}
\IEEEauthorblockA{\textit{Federal University of Ceará}\\
Fortaleza, Brazil \\
darlan@gtel.ufc.br}
\and
\hspace{50pt}
\IEEEauthorblockN{Tarcisio F. Maciel}
\IEEEauthorblockA{\hspace{50pt}\textit{Federal University of Ceará}\\
\hspace{50pt}
Fortaleza, Brazil \\
\hspace{50pt}
maciel@gtel.ufc.br}
\and
\IEEEauthorblockN{Fco. Rafael M. Lima}
\IEEEauthorblockA{\textit{Federal University of Ceará}\\
Sobral, Brazil \\
rafaelm@gtel.ufc.br}
\and
\IEEEauthorblockN{Behrooz Makki}
\IEEEauthorblockA{\textit{Ericsson Research}\\
Gothenburg, Sweden \\
behrooz.makki@ericsson.com}
}

\maketitle

\begin{abstract}
To address the need of coverage enhancement in the \ac{5G} of wireless cellular telecommunications, while taking into account possible bottlenecks related to deploying fiber based backhaul (e.g., required cost and time), the \ac{3GPP} proposed in Release 18 the concept of \acp{NCR}. %
\acp{NCR} enhance previous \ac{RF} repeaters by exploring beamforming transmissions controlled by the network through side control information. %
In this context, this paper introduces the concept of \ac{NCR}. %
Furthermore, we present a system level model that allows the performance evaluation of an \ac{NCR}-assisted network. %
Finally, we evaluate the network deployment impact on the performance of \ac{NCR}-assisted networks. %
As we show, with proper network planning, \acp{NCR} can boost the \ac{SINR} of the \acp{UE} in a poor coverage of a macro base station. %
Furthermore, cell-edge \acp{UE} and \ac{UL} communications are the ones that benefit the most from the presence of \acp{NCR}. %
%
\end{abstract}

\acresetall

\section{Introduction} 
\label{SEC:Intro}

The \ac{5G} of the wireless cellular telecommunications systems was initially standardized in Release 15 of the \ac{3GPP} and continued in the next releases. %
One of its advances was the incorporation of carrier frequencies up to \SI{52.6}{GHz}, i.e., \ac{mmWave} bands. %
Some of the advantages of using \ac{mmWave} are the large spectrum available at these frequencies, the possibility of transmitting with narrow beams and, as a consequence, the capacity of achieving throughput higher than previous generations~\cite{Rangan2014}. %

Deploying a system in \ac{mmWave} is challenging due to its smaller coverage range compared with previous generations, which work on sub-\SI{6}{GHz} band. %
The main reason for this is the more aggressive path loss and higher obstacle attenuation at higher frequencies~\cite{Rangan2014}. %

One envisioned solution to overcome the problem of \ac{mmWave} small coverage area is to increase network densification, i.e., increasing the number of network nodes that can provide access to a \ac{UE}~\cite{Jungnickel2014}. %
These network nodes can be connected to the \ac{CN} either via fiber or via a wireless connection. %

Although fiber provides reliable high peak rate, one of the drawbacks of fiber backhaul is its high cost and time to be deployed by virtue of trenching and installation. %
Furthermore, depending on the region, such as historic and preservation areas, it can be impossible to install the fiber. %
Hence, wireless links have been considered as an alternative solution to backhaul instead of fiber~\cite{Madapatha2020, Madapatha2021, Dang2020}. %
Some of the new network nodes with wireless backhaul that have been considered for \ac{5G} are: \ac{NCR}~\cite{Carvalho2024}, \ac{IAB} node~\cite{Monteiro2022} and \ac{RIS}~\cite{Astrom2024}. %

In previous generations, \ac{RF} repeaters were used to relay signals extending the system coverage. %
An \ac{RF} repeater is a non-regenerative node that amplifies a received signal. %
The main advantages of \ac{RF} repeaters are their low cost, easy implementation, and small impact in system latency. %
Nevertheless, they have at least one main disadvantage: the omnidirectional transmissions. %
So, recently, \ac{3GPP} studied, in Release 18~\cite{3gpp.38.867}, an enhanced \ac{RF} repeater, called \ac{NCR}. %
\Acp{NCR} enhance previous \ac{RF} repeaters by exploring beamforming transmissions, which are controlled by the network through side control information~\cite{Sousa2024}. 

In this context, this paper introduces the concept of \ac{NCR}. %
Furthermore, we present a system level model that allows the performance evaluation of an \ac{NCR}-assisted network. %
Finally, we evaluate the impact of four \ac{NCR} deployment options on the performance of \ac{NCR}-assisted networks. %

This paper is organized as follows. %
\SecRef{SEC:System_Model} presents the system model. %
The considered scenarios are described in \SecRef{SEC:NCR_Deployment}. %
\SecRef{SEC:Perf_Eval} presents our simulation setup, computational results and discussions. %
Finally, \SecRef{SEC:Conclusions} summarizes our main findings and presents our future perspectives. %


\section{System Model}
\label{SEC:System_Model}

\begin{figure}[t]
	\centering
	\includegraphics[width=0.4\columnwidth]{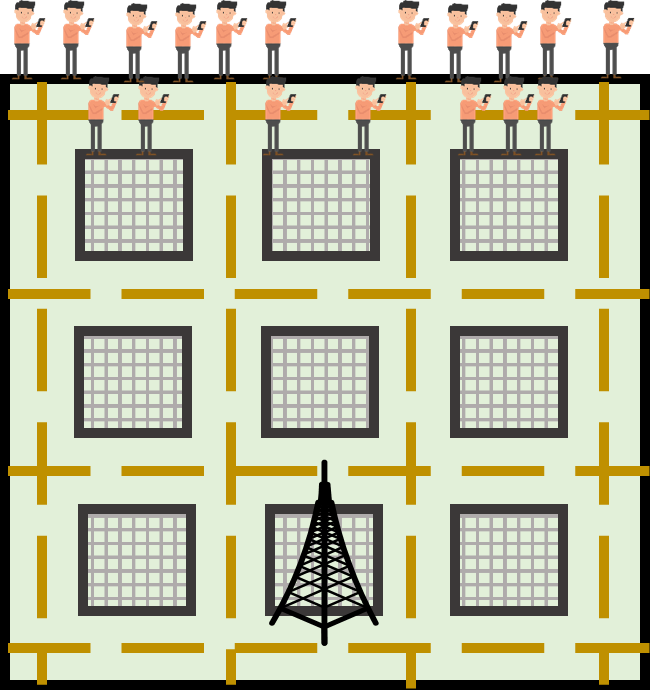}
	\caption{Benchmark scenario with a crowded street on the top and a macro gNB in the middle bottom block.}
	\label{FIG:only_macro}
\end{figure}

Consider a Madrid grid based scenario~\cite{METIS:D6.1:2013}, as illustrated in \FigRef{FIG:only_macro}, where there is a crowded street on the top and a macro \ac{gNB} in the middle bottom block. %
This represents a daily situation of urban areas where there are a lot of people and the closest \ac{gNB} is some blocks away. %
In order to enhance the system coverage area, \acp{NCR} are deployed in specific positions that will be presented later. %

\begin{figure}[t]
	\centering
	\includegraphics[width=1.0\columnwidth]{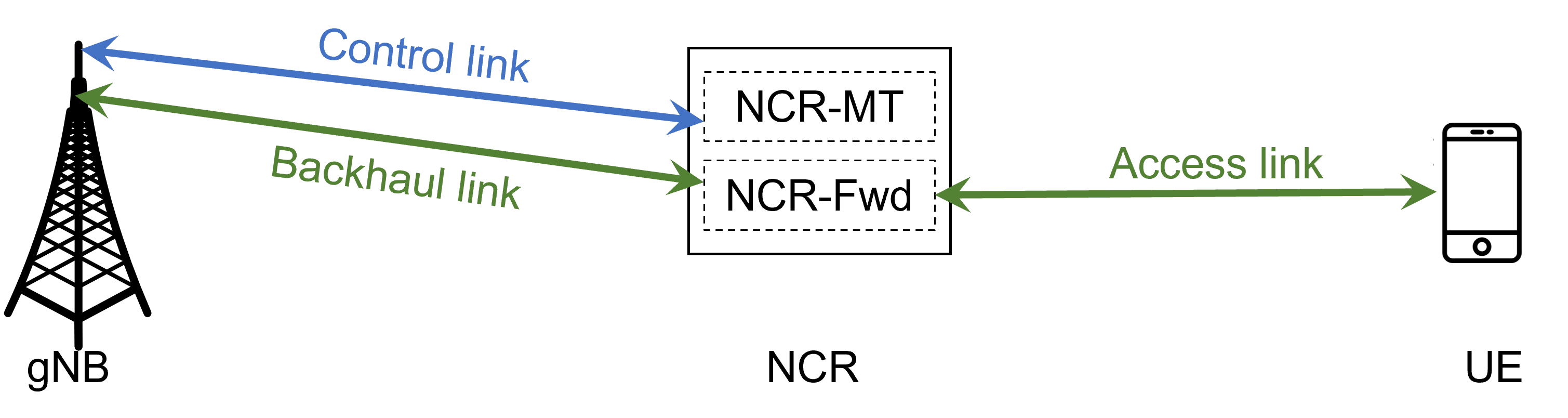}
	\caption{NCR links.}
	\label{FIG:ncr_model}
\end{figure}

An \ac{NCR} is deployed and under the control of a mobile network operator and, for all management purposes, is logically part of its controlling \ac{gNB}. %
This enables to extend the coverage and cover the coverage holes where the \ac{gNB} alone may not be able to cover. %
\ac{NCR} can be split into two parts: \ac{NCR}-\ac{MT} and \ac{NCR}-\ac{Fwd}, as shown in  \FigRef{FIG:ncr_model}. %
The first one is responsible for exchanging control information with the controlling \ac{gNB} via the control link, based on Uu interface, that exchanges  the side control information between \ac{gNB} and \ac{NCR}. %
The second one is responsible for the \ac{AF} relaying in access link and backhaul link, being controlled by the side control information received by the \ac{NCR}-\ac{MT}. %

In our considered model, transmissions of control and backhaul links are performed at the \ac{NCR} by a specific panel, called herein backhaul panel, while transmission of access links are performed by panels called herein access panels. %
We remark that control and backhaul links could be deployed on different panels. %
Since \ac{gNB} and \ac{NCR} are fixed, the backhaul and control links are well planned with strong \ac{LOS} connection steered toward the \ac{gNB}. %
Regarding access links, NCR can deploy more than one access panel. %
One of the advantages of using more than one access panel is to increase the coverage area. %
One strategy to mitigate interference between the access and backhaul panels is, e.g., to deploy them with angular difference higher than or equal to $120$~degrees~\cite{Dong2023}. %

\begin{figure}[t]
	\centering
	\includegraphics[width=1.0\columnwidth]{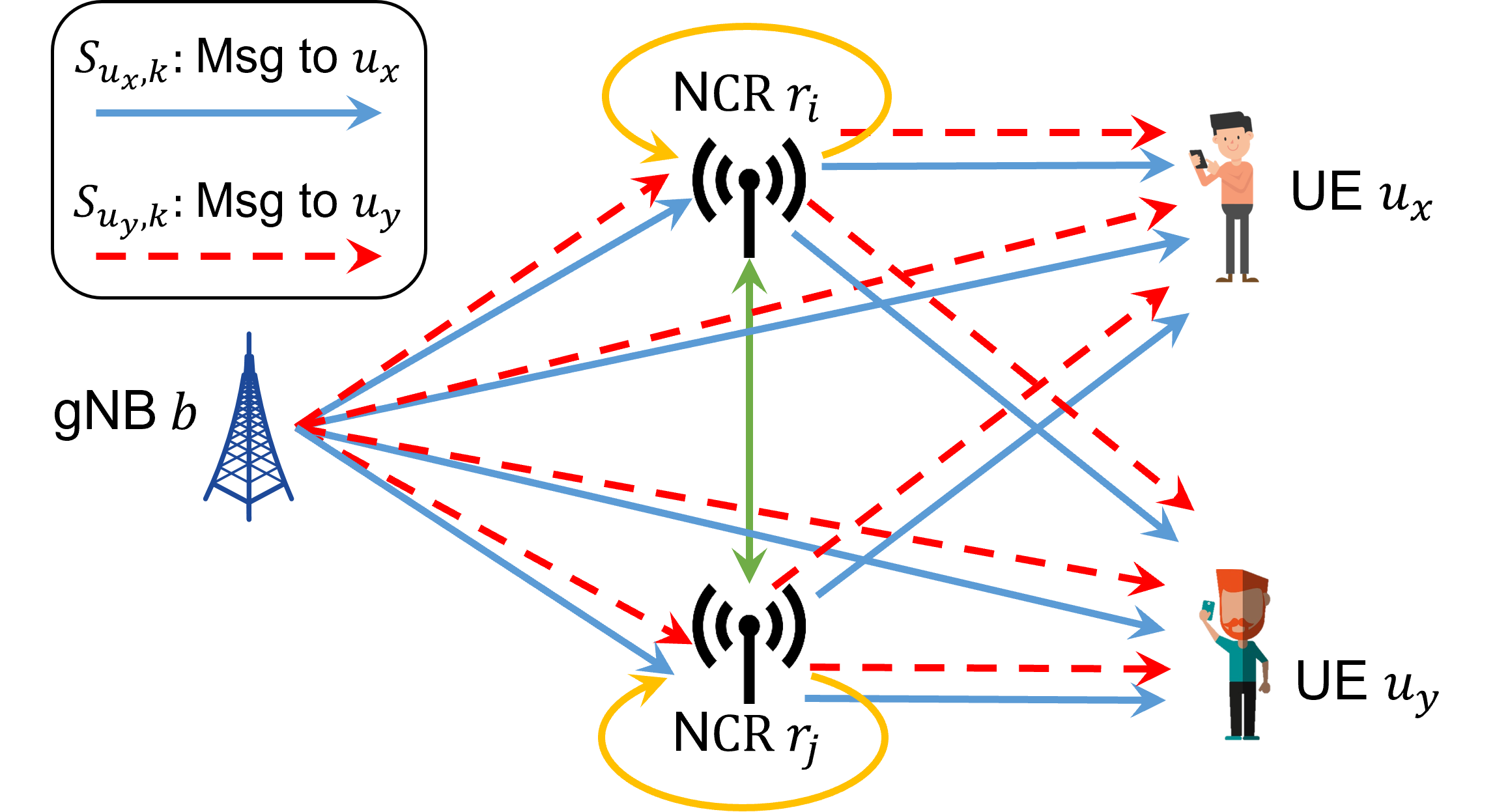}
	\caption{Links considered in the SINR estimation.}
	\label{FIG:sinr}
\end{figure}

The management of links between \ac{gNB}-\ac{NCR} and \ac{NCR}-\ac{UE} is performed by the \ac{gNB}. %
For this, two procedures of beam sweeping are performed, one for each link. %
The  beam sweeping for the \ac{gNB}-\ac{NCR} and \ac{NCR}-\ac{UE} links are performed at every $t_{backhaul}$ and $t_{access}$ instants of time, respectively. %
The \ac{gNB}-\ac{NCR} link has higher coherence time than the \ac{NCR}-\ac{UE} link, since \ac{gNB} and \ac{NCR} are fixed while the \acp{UE} move around. %
Thus, $t_{backhaul}$ can have considerably higher values than $t_{access}$. %
The choice of whether a \ac{UE} is going to be directly served by the \ac{gNB} or through the \ac{NCR} can be summarized as follows: a \ac{UE} measures the quality of the beams that it can hear from the \ac{gNB} and \ac{NCR}; after that it transmits a measurement report to the \ac{gNB}, which can go through the \ac{NCR} if the \ac{UE} is already attached to it; based on the received measurement reports the \ac{gNB} decides whether the \ac{UE} is going to be directly connected to it or if the connection is going to pass through the \ac{NCR} and also which beams are going to be used for the communication. %
It is important to note that, according to \ac{3GPP} Release~18, \ac{NCR} is transparent to the\ac{UE}, i.e., the \ac{UE} does not understand the presence of the \ac{NCR} and it follows the same behavior as in the cases with no \acp{NCR}. %

Regarding the estimation of \ac{SINR} perceived by a \ac{UE} in the \ac{DL}, it is calculated as follows. %
Consider that a \ac{RB} is the minimum allocable frequency block, where one \ac{RB} is composed of a number of adjacent subcarriers in the frequency domain. %
The bandwidth is split into $K$ \acp{RB}. %
As illustrated in \FigRef{FIG:sinr}, also consider that \ac{gNB} $b$ controls \acp{NCR} $r_{i}$ and $r_{j}$. %
Moreover, consider that \acp{UE} $u_{x}$ and $u_{y}$ are connected to \ac{gNB} $b$ through \acp{NCR} $r_{i}$ and $r_{j}$, respectively. %
Furthermore, consider that a \ac{NCR} does not perform signal processing, i.e., it neither filters incoming signals that are not addressed to a \ac{UE} not served by itself nor filters out signals coming from \acp{gNB} different from the one that they are connected to. %
Thus, the useful power $S_{u_{x},k}$ received by \ac{UE} $u_{x}$ at \ac{RB} $k$ can be expressed as:

\begin{equation}
	\begin{aligned}
		S_{u_{x},k} &= \gamma_{b,u_{x},k} \cdot p_{b,u_{x},k} \\
		& \hspace{15pt} + \sum_{r_{i}\in \stR}\left(\gamma_{r_{i},u_{x},k} \cdot g_{r_{i},k} \cdot \gamma_{b,r_{i},k} \cdot p_{b,u_{x},k}\right),
	\end{aligned}
\end{equation}
\noindent where $\gamma_{i,j,k} = \textbf{d}_{j,k} \textbf{H}_{i,j,k} \textbf{f}_{i,k}$ denotes the combined effect of the channel $\textbf{H}_{i,j,k}$ after the transmission and reception filters $\textbf{f}_{i,k}$ and $\textbf{d}_{j,k}$, respectively, applied to a signal transmitted from a node $i$ to a node $j$; $p_{b,u_{x},k}$ is the transmit power used by \ac{gNB} $b$ to transmit useful signal at \ac{RB} $k$; $g_{r_{i},k}$ is the power gain applied by \ac{NCR} $r_{i}$ at \ac{RB} $k$; and $\stR$ is the set of \acp{NCR}. %

The signal sent by \ac{gNB} $b$ to \ac{UE} $u_{y}$ at \ac{RB} $k$ is seen by \ac{UE} $u_{x}$ as interference. %
Thus, the interference $I_{u_{x},k}$ perceived by \ac{UE} $u_{x}$ at \ac{RB} $k$ is equal to:
\begin{equation}
	\begin{aligned}
		I_{u_{x},k} &= \sum_{u_{y} \in \stU, y\neq x} S_{u_{y},k} \hspace{10pt} \\
		& = \sum_{u_{y} \in \stU, y\neq x} \bigg[\gamma_{b,u_{y},k} \cdot p_{b,u_{y},k} \\
		& \hspace{10pt} + \sum_{r_{i}\in \stR}\left(\gamma_{r_{i},u_{y},k} \cdot g_{r_{i},k} \cdot \gamma_{b,r_{i},k} \cdot p_{b,u_{y},k}\right) \bigg],
	\end{aligned} 
\end{equation}

\noindent where $\stU$ is the set of \acp{UE} in the system. %

Neither the interference between \acp{NCR}, i.e., green arrow in \FigRef{FIG:sinr}, nor the \ac{NCR} self-interference, i.e., yellow arrows in \FigRef{FIG:sinr}, are considered in our simulation setup, %
It is assumed that both of them can be mitigated in the network deployment phase. %
Specifically, the first one can be mitigated ensuring that a \ac{NCR} is not in the coverage area of another \ac{NCR}. %
As already mentioned, the second one can be mitigated ensuring that the \ac{NCR} panels have an angular difference higher than or equal to 120 degrees~\cite{Dong2023}. %

The overall noise power $N_{u_{x},k}$ received by \ac{UE} $u_{x}$ at \ac{RB} $k$ is the \ac{AWGN} added of the noise components amplified by the \acp{NCR}. %
It can be expressed as:
\begin{equation}
	N_{u_{x},k} = \sigma^{2}_{k} \left[1 + \sum_{r_{i} \in \stR} \gamma_{r_{i},u_{x},k} \cdot g_{r_{i},k} \right].
\end{equation}

Finally, the \ac{SINR} $\rho_{u_{x},k}$ perceived by \ac{UE} $u_{x}$ at \ac{RB} $k$ is 
\begin{equation}
	\rho_{u_{x},k} = \frac{S_{u_{x},k}}{I_{u_{x},k} + N_{u_{x},k}}.
\end{equation}


\section{\Ac{NCR} Deployment}
\label{SEC:NCR_Deployment}

In this section, we present the benchmark scenario plus the four \ac{NCR} considered scenarios, each one with different \ac{NCR} characteristics. %
The first scenario, the benchmark, is the one presented in \FigRef{FIG:only_macro} with only one \ac{gNB} serving the \acp{UE}. %
Aiming at improving the benchmark's system performance, we deploy \ac{NCR} nodes in the scenario of \FigRef{FIG:only_macro}. %
Different possibilities are considered. %

\begin{figure}[t]
	\centering
	
	\subfloat{%
		\includegraphics[width=0.8\columnwidth]{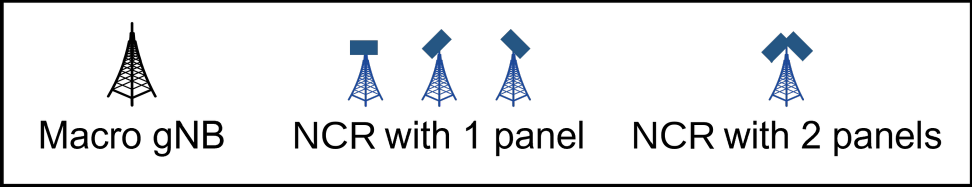}
	}
	
	\setcounter{subfigure}{0}

	\subfloat[Scenario 2: 1 \acs{NCR} with 1 panel.]{
		\includegraphics[width=0.35\columnwidth]{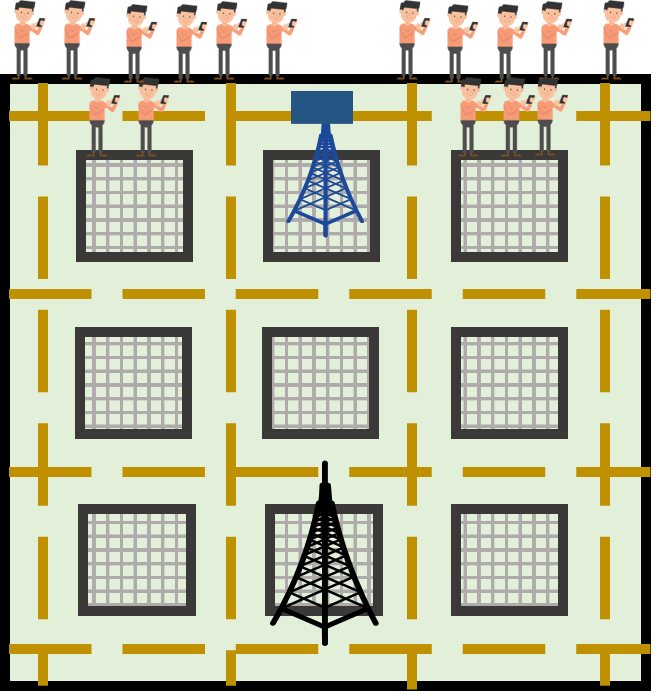}
		\label{FIG:ncr1panel1}
	}
	\hspace{0.1pt}
	\subfloat[Scenario 3: 1 \acs{NCR} with 2 panels.]{
		\includegraphics[width=0.35\columnwidth]{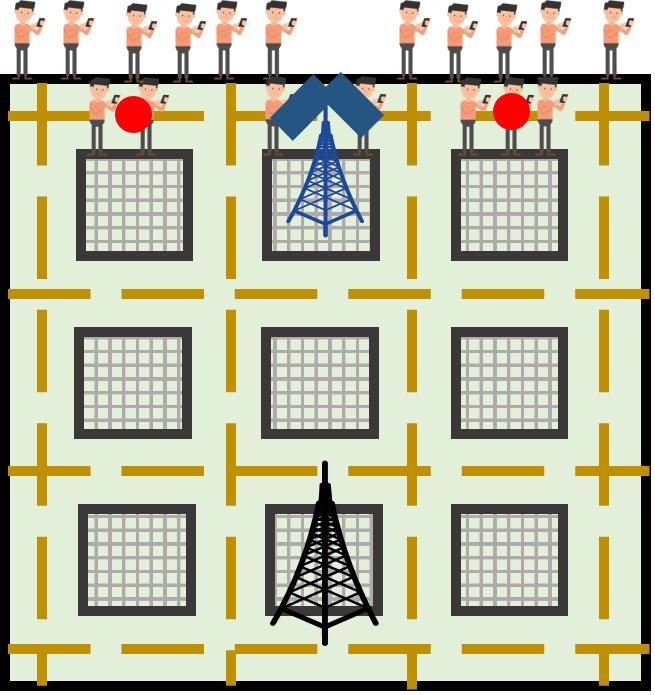}
		\label{FIG:ncr1panel2}
	}
	\vspace{1pt}
	\subfloat[Scenario 4: 2 \acp{NCR} with 1 panel each.]{
		\includegraphics[width=0.35\columnwidth]{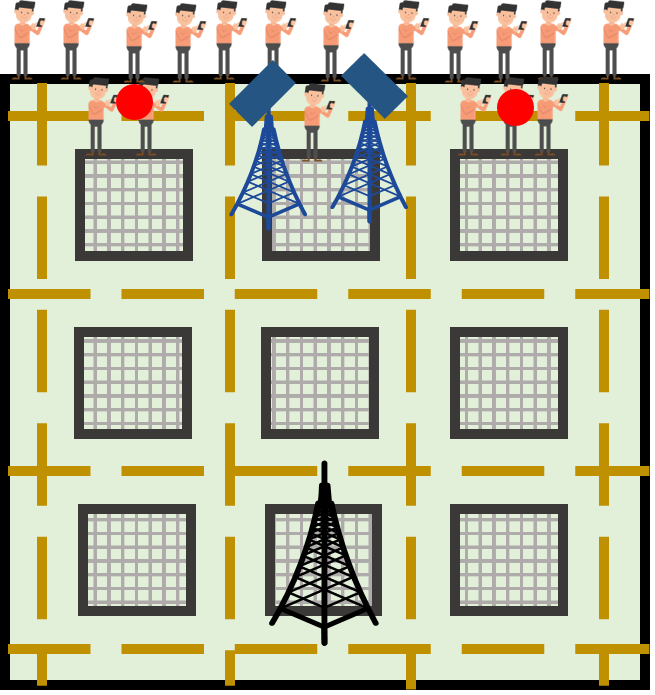}
		\label{FIG:ncr2panel1}
	}
	\hspace{0.1pt}
	\subfloat[Scenario 5: 2 \acs{NCR} with 1 panel each, deployed at the side blocks.]{
		\includegraphics[width=0.35\columnwidth]{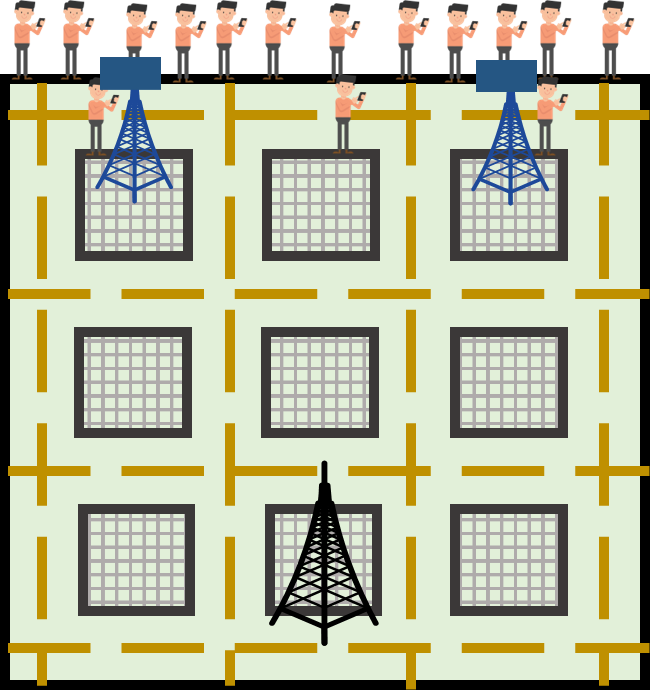}
		\label{FIG:ncr2panel1-edge}
	}
	
	\caption{Evaluated scenarios.}
\end{figure}

In the second scenario, one \ac{NCR} node is positioned in the middle top block with one access panel steering towards the middle of the street in front of its block, as shown in \FigRef{FIG:ncr1panel1}. %
The third scenario, \FigRef{FIG:ncr1panel2}, has two access panels, instead of only one as in the second scenario, and they point towards the middle of the street in front of their side blocks represented by the red dots in \FigRef{FIG:ncr1panel2}. %
So, the difference between scenarios 2 and 3 is the amount of access panels and their steering direction. %
In the fourth scenario, \FigRef{FIG:ncr2panel1}, there are two \ac{NCR} nodes with only one access panel, each one positioned in one corner of the middle top block and also pointing towards the middle of the street in front of their side blocks (red dots). %
The fifth scenario is similar to the fourth with the only difference being the \ac{NCR} nodes' position, as illustrated in \FigRef{FIG:ncr2panel1-edge}. %

These different scenarios create the possibility to analyze the \ac{NCR}'s performance from three perspectives: the number of \acp{NCR}, their position and the number of antenna arrays available for the access links in each \ac{NCR}. %

\section{Performance Evaluation}
\label{SEC:Perf_Eval}

In this section, we present a comparison between the performance of the five scenarios described in the previous section based on computational simulations. %
Section \ref{SEC:Sim_Assumptions} presents the details concerning the considered simulation modeling and Section \ref{SEC:Sim_Results} discusses the simulation results. %

\begin{table*}[!t] 
	\centering
	\small
	\caption{TDD scheme adopted.}
	\label{TABLE:macro-TDD}
	\resizebox{\textwidth}{!}{
		\begin{tabular}{l|llllllllll|ll|l}
			\toprule
			Slot                & 1  & 2      & 3  & 4  & 5  & 6  & 7      & 8  & 9  & 10 & DL usage & UL usage & Total usage \\ 
			\midrule
			Macro and Pico gNBs & DL & S (DL) & UL & UL & UL & DL & S (DL) & UL & UL & DL & 50\%     & 50\%     & 100\% \\
			\bottomrule
		\end{tabular}
	}
\end{table*}

\begin{table*}[!t]
	\centering
	\caption{Entities characteristics.}
	\label{TABLE:Entities-characteristics}
	\begin{tabularx}{\textwidth}{lXXX}
		\toprule
		\textbf{Parameter} & \textbf{Macro \ac{gNB}} & \textbf{\ac{NCR}} & \textbf{\ac{UE}} \\
		\midrule
		Height & \SI{25}{\meter} & \SI{10}{\meter} & \SI{1.5}{\meter} \\
		Transmit power & \SI{35}{\decibel m} & \SI{33}{\decibel m} & \SI{24}{\decibel m} \\
		Antenna array & URA $8\times 8$ & URA $8\times 8$ (2 panels) & Single Antenna \\
		Antenna tilt & $12^{\circ}$ & Backhaul panel: Aligned with \ac{gNB}; Access panel: $12^{\circ}$ & $0^{\circ}$ \\
		Antenna element pattern & \ac{3GPP} 3D~\cite{3gpp.38.901} & \ac{3GPP} 3D~\cite{3gpp.38.901} & Omni \\
		Max. antenna element gain & \SI{8}{\decibel i} & \SI{8}{\decibel i} & \SI{0}{\decibel i} \\
		Speed & \SI{0}{km/h} & \SI{0}{km/h}  & \SI{3}{km/h} \\
		\bottomrule
	\end{tabularx}
\end{table*}

\begin{table}
	\centering
	\setlength{\tabcolsep}{1ex}
	\caption{Simulation parameters.}
	\label{TABLE:Simul_Param}
	\begin{tabularx}{0.99\columnwidth}{>{\raggedright\arraybackslash}X>{\raggedright\arraybackslash}X}
		\toprule
		\textbf{Parameter} & \textbf{Value} \\
		\midrule
		Carrier frequency & \SI{28}{\GHz}\\
		System bandwidth & \SI{50}{\MHz}\\
		Subcarrier spacing & \SI{60}{\kHz}\\
		Number of subcarriers per \acs{RB} &  $12$\\
		Number of \acsp{RB} & $66$\\
		Slot duration & \SI{0.25}{\ms} \\
		OFDM symbols per slot & $14$ \\
		Channel generation procedure & As described in~\cite[Fig.7.6.4-1]{3gpp.38.901}\\
		Path loss  & Eqs. in~\cite[Table 7.4.1-1]{3gpp.38.901}\\
		Fast fading & As described in~\cite[Sec.7.5]{3gpp.38.901} and \cite[Table 7.5-6]{3gpp.38.901} \\
		AWGN density power per subcarrier & \SI{-174}{dBm/Hz}\\
		Noise figure &  \SI{9}{\decibel}\\
		Number of \acp{UE} & 72 \\
		\acs{CBR} packet size & $3072$ bits \\
		\bottomrule
	\end{tabularx}
\end{table}
\subsection{Simulation Assumptions}
\label{SEC:Sim_Assumptions}

In the Madrid grid, it is considered nine \SI{120}{m} $\times$ \SI{120}{m} blocks surrounded by \SI{3}{m} wide sidewalks. %
The street's width is \SI{14}{m}. %
The \acp{UE} are randomly positioned, following a uniform distribution, in both sidewalks on the top street in all scenarios. %
The \acp{UE} are allowed to walk only in the sidewalks of this street. %

The considered channel model, described in~\cite{Pessoa2019}, is an implementation of the \ac{3GPP} channel model standardized in~\cite{3gpp.38.901}. %
It is spatially and time consistent. %
Furthermore, it considers a  distance-dependent path-loss, a lognormal shadowing component, and a small-scale fading. %

In the time domain, a slot, the minimum scheduling unit, has a duration of \SI{0.25}{ms} and is composed of 14 \ac{OFDM} symbols. %
\TabRef{TABLE:macro-TDD} presents the adopted \ac{TDD} scheme, which is standardized by \ac{3GPP} in \cite{3gpp.36.211b}. %

In the frequency domain, one \ac{RB} consists of $12$~consecutive subcarriers with a subcarrier spacing of \SI{60}{kHz}. %
The transmissions are performed at \SI{28}{GHz} in a bandwidth of \SI{50}{MHz}. %
The \ac{RR} scheduler is adopted to schedule the \acp{RB}. %
The \ac{RR} iteratively allocates the \acp{RB}, scheduling in a given \ac{RB} the \ac{UE} bearer waiting the longest time in the queue. %
The \ac{RR} was chosen since it is a well known scheduler enabling an interested reader to reproduce our performance evaluation. %
Besides, our main objective is not to find the scheduler that optimizes the system behavior, but rather compare the different solutions under the same conditions, i.e., using the same scheduler. %

The \ac{CQI}/\ac{MCS} mapping curves standardized in \cite{3gpp.38.214} are used with a target \ac{BLER} of \SI{10}{\%}. %
To avoid the increase of the \ac{BLER}, it is considered an outer loop strategy, i.e, when a transmission error occurs, the estimated \ac{SINR} decreases \SI{1}{dB}, and, when a transmission error does not occur, the estimated \ac{SINR} increases \SI{0.1}{dB}. %
\TabRef{TABLE:Entities-characteristics} and \TabRef{TABLE:Simul_Param} summarize the most relevant simulation parameters. %

\subsection{Simulation Results}
\label{SEC:Sim_Results}

\begin{figure}[t]
	\centering
	\subfloat[\acs{CDF} of \acs{SINR} of all \acp{UE}.]{
			\begin{tikzpicture}
		\begin{axis}[common plots axis options,
			ylabel=CDF,
			xlabel=SINR (dB),
			xmin = -10, xmax = 45,
			xtick = {-10,-5,...,45},
			legend style={
		    	at = {(0.5, 1.05)},
		    	anchor = south,
		    	legend columns = 2,
			}
			]
			\pgfplotstableread [col sep=comma] {\plotsDataPath/sinr_per_block_downlink.csv}\tableData
			
			\addplot[scenario0 style]
			table[x=all_NCR_0_x, y=all_NCR_0_y] from \tableData;
			\addlegendentry{Only Macro}
			
			\addplot[scenario1 style]
			table[x=all_NCR_1_x, y=all_NCR_1_y] from \tableData;
			\addlegendentry{1 NCR}
			
			\addplot[scenario12 style]
			table[x=all_NCR_1_2_x, y=all_NCR_1_2_y] from \tableData;
			\addlegendentry{1 NCR w/ 2 panels}
			
			\addplot[scenario2 style]
			table[x=all_NCR_2_x, y=all_NCR_2_y] from \tableData;
			\addlegendentry{2 NCRs @ center block corners}
			
			\addplot[scenario21 style]
			table[x=all_NCR_2_1_x, y=all_NCR_2_1_y] from \tableData;
			\addlegendentry{2 NCRs @ side blocks}
			
		\end{axis}
	\end{tikzpicture}
	
		\label{FIG:Simulation-results-ncr-sinr-all-downlink}
	}	
	
	\subfloat[\acs{CDF} of \acs{SINR} of \acp{UE} in the central block.]{
			\begin{tikzpicture}
		\begin{axis}[common plots axis options,
			ylabel=CDF,
			xlabel=SINR (dB),
			xmin = -10, xmax = 45,
			xtick = {-10,-5,...,45},
			legend style={
		    	at = {(0.5, 1.05)},
		    	anchor = south,
		    	legend columns = 3,
			}
			]
			\pgfplotstableread [col sep=comma] {\plotsDataPath/sinr_per_block_downlink.csv}\tableData
			
			\addplot[scenario0 style]
			table[x=Q2_NCR_0_x, y=Q2_NCR_0_y] from \tableData;
			
			\addplot[scenario1 style]
			table[x=Q2_NCR_1_x, y=Q2_NCR_1_y] from \tableData;
			
			\addplot[scenario12 style]
			table[x=Q2_NCR_1_2_x, y=Q2_NCR_1_2_y] from \tableData;
			
			\addplot[scenario2 style]
			table[x=Q2_NCR_2_x, y=Q2_NCR_2_y] from \tableData;
			
			\addplot[scenario21 style]
			table[x=Q2_NCR_2_1_x, y=Q2_NCR_2_1_y] from \tableData;
			
		\end{axis}
	\end{tikzpicture}
	
		\label{FIG:Simulation-results-ncr-sinr-q2-downlink}
	}	
	
	\subfloat[CDF of SINR of UEs in the side block.]{
			\begin{tikzpicture}
		\begin{axis}[common plots axis options,
			ylabel=CDF,
			xlabel=SINR (dB),
			xmin = -10, xmax = 45,
			xtick = {-10,-5,...,45},
			legend style={
		    	at = {(0.5, 1.05)},
		    	anchor = south,
		    	legend columns = 3,
			}
			]
			\pgfplotstableread [col sep=comma] {\plotsDataPath/sinr_per_block_downlink.csv}\tableData
			
			\addplot[scenario0 style]
			table[x=Q3_NCR_0_x, y=Q3_NCR_0_y] from \tableData;
			
			\addplot[scenario1 style]
			table[x=Q3_NCR_1_x, y=Q3_NCR_1_y] from \tableData;
			
			\addplot[scenario12 style]
			table[x=Q3_NCR_1_2_x, y=Q3_NCR_1_2_y] from \tableData;
			
			\addplot[scenario2 style]
			table[x=Q3_NCR_2_x, y=Q3_NCR_2_y] from \tableData;
			
			\addplot[scenario21 style]
			table[x=Q3_NCR_2_1_x, y=Q3_NCR_2_1_y] from \tableData;
			
		\end{axis}
	\end{tikzpicture}
	
		\label{FIG:Simulation-results-ncr-sinr-q3-downlink}
	}	
	\caption{CDF of SINR in downlink. }
	\label{FIG:Simulation-results-ncr-sinr-q2-q3-downlink}
\end{figure}
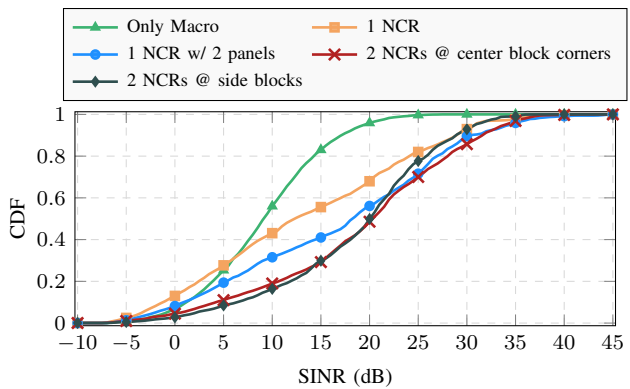
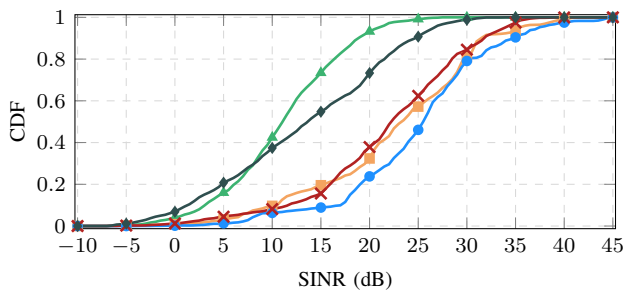
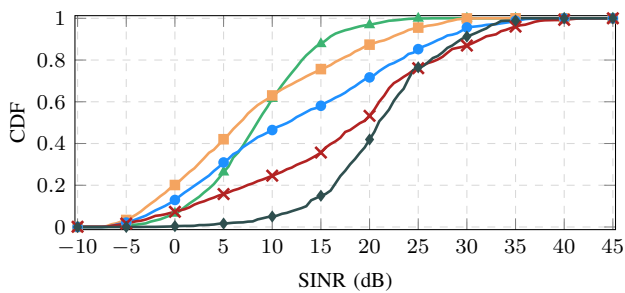
\begin{figure}[t]
	\centering
	\subfloat[\acs{CDF} of \acs{SINR} of all \acp{UE}.]{
			\begin{tikzpicture}
		\begin{axis}[common plots axis options,
			ylabel=CDF,
			xlabel=SINR (dB),
			xmin = -10, xmax = 45,
			xtick = {-10,-5,...,45},
			legend style={
		    	at = {(0.5, 1.05)},
		    	anchor = south,
		    	legend columns = 2,
			}
			]
			\pgfplotstableread [col sep=comma] {\plotsDataPath/sinr_per_block_uplink.csv}\tableData
			
			\addplot[scenario0 style]
			table[x=all_NCR_0_x, y=all_NCR_0_y] from \tableData;
			\addlegendentry{Only Macro}
			
			\addplot[scenario1 style]
			table[x=all_NCR_1_x, y=all_NCR_1_y] from \tableData;
			\addlegendentry{1 NCR}
			
			\addplot[scenario12 style]
			table[x=all_NCR_1_2_x, y=all_NCR_1_2_y] from \tableData;
			\addlegendentry{1 NCR w/ 2 panels}
			
			\addplot[scenario2 style]
			table[x=all_NCR_2_x, y=all_NCR_2_y] from \tableData;
			\addlegendentry{2 NCRs @ center block corners}
			
			\addplot[scenario21 style]
			table[x=all_NCR_2_1_x, y=all_NCR_2_1_y] from \tableData;
			\addlegendentry{2 NCRs @ side blocks}
			
		\end{axis}
	\end{tikzpicture}
	
		\label{FIG:Simulation-results-ncr-sinr-all-uplink}
	}	
	
	\subfloat[\acs{CDF} of \acs{SINR} of \acp{UE} in the central block.]{
			\begin{tikzpicture}
		\begin{axis}[common plots axis options,
			ylabel=CDF,
			xlabel=SINR (dB),
			xmin = -10, xmax = 45,
			xtick = {-10,-5,...,45},
			legend style={
		    	at = {(0.5, 1.05)},
		    	anchor = south,
		    	legend columns = 3,
			}
			]
			\pgfplotstableread [col sep=comma] {\plotsDataPath/sinr_per_block_uplink.csv}\tableData
			
			\addplot[scenario0 style]
			table[x=Q2_NCR_0_x, y=Q2_NCR_0_y] from \tableData;
			
			\addplot[scenario1 style]
			table[x=Q2_NCR_1_x, y=Q2_NCR_1_y] from \tableData;
			
			\addplot[scenario12 style]
			table[x=Q2_NCR_1_2_x, y=Q2_NCR_1_2_y] from \tableData;
			
			\addplot[scenario2 style]
			table[x=Q2_NCR_2_x, y=Q2_NCR_2_y] from \tableData;
			
			\addplot[scenario21 style]
			table[x=Q2_NCR_2_1_x, y=Q2_NCR_2_1_y] from \tableData;
			
		\end{axis}
	\end{tikzpicture}
	
		\label{FIG:Simulation-results-ncr-sinr-q2-uplink}
	}	
	
	\subfloat[CDF of SINR of UEs in the side block.]{
			\begin{tikzpicture}
		\begin{axis}[common plots axis options,
			ylabel=CDF,
			xlabel=SINR (dB),
			xmin = -10, xmax = 45,
			xtick = {-10,-5,...,45},
			legend style={
		    	at = {(0.5, 1.05)},
		    	anchor = south,
		    	legend columns = 3,
			}
			]
			\pgfplotstableread [col sep=comma] {\plotsDataPath/sinr_per_block_uplink.csv}\tableData
			
			\addplot[scenario0 style]
			table[x=Q3_NCR_0_x, y=Q3_NCR_0_y] from \tableData;
			
			\addplot[scenario1 style]
			table[x=Q3_NCR_1_x, y=Q3_NCR_1_y] from \tableData;
			
			\addplot[scenario12 style]
			table[x=Q3_NCR_1_2_x, y=Q3_NCR_1_2_y] from \tableData;
			
			\addplot[scenario2 style]
			table[x=Q3_NCR_2_x, y=Q3_NCR_2_y] from \tableData;
			
			\addplot[scenario21 style]
			table[x=Q3_NCR_2_1_x, y=Q3_NCR_2_1_y] from \tableData;
			
		\end{axis}
	\end{tikzpicture}
	
		\label{FIG:Simulation-results-ncr-sinr-q3-uplink}
	}	
	\caption{CDF of SINR in uplink. }
	\label{FIG:Simulation-results-ncr-sinr-q2-q3-uplink}
\end{figure}
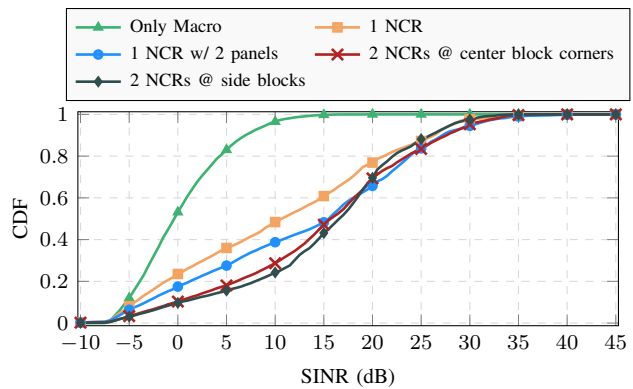
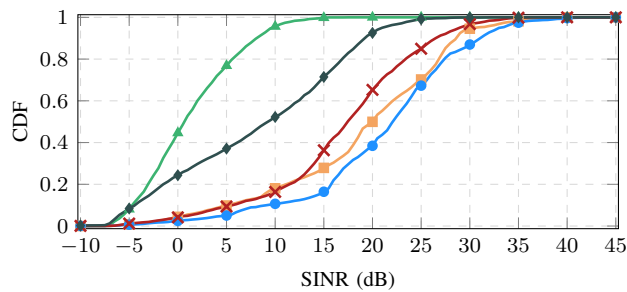
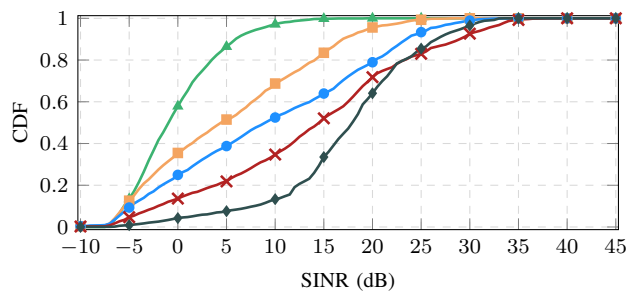

\FigRef{FIG:Simulation-results-ncr-sinr-q2-q3-downlink} and \FigRef{FIG:Simulation-results-ncr-sinr-q2-q3-uplink} present the \ac{CDF} of the \acp{UE} \ac{SINR} in \ac{DL} and \ac{UL}, respectively. 
More specifically, \FigRef{FIG:Simulation-results-ncr-sinr-all-downlink} and \FigRef{FIG:Simulation-results-ncr-sinr-all-uplink} consider all the \acp{UE}, while \FigRef{FIG:Simulation-results-ncr-sinr-q2-downlink} and \FigRef{FIG:Simulation-results-ncr-sinr-q2-uplink} consider only the \acp{UE} in the central block, and \FigRef{FIG:Simulation-results-ncr-sinr-q3-downlink} and \FigRef{FIG:Simulation-results-ncr-sinr-q3-uplink} consider only the \acp{UE} in the side blocks. %

First of all, as expected, we can see that compared to the benchmark solution with no \ac{NCR}, deploying \acp{NCR} improves the \acp{UE} \ac{SINR} in both \ac{DL} and \ac{UL}. %
The \acp{UE} in the side blocks correspond, in average, to two thirds of the total number of \acp{UE} and, in the benchmark scenario, they have lower values of \ac{SINR} than the \acp{UE} in the central block. %
This is mainly due to their higher distance to the \ac{gNB}. %
Thus, solutions that improve the performance of the \acp{UE} in the side blocks have a higher gain in the lower part of the \acp{CDF} presented in \FigRef{FIG:Simulation-results-ncr-sinr-q2-q3-downlink} and \FigRef{FIG:Simulation-results-ncr-sinr-q2-q3-uplink}. %
More specifically, deploying \acp{NCR} in the side blocks is the solution that best improves side blocks \acp{UE} \ac{SINR} (\FigRef{FIG:Simulation-results-ncr-sinr-q3-downlink} and \FigRef{FIG:Simulation-results-ncr-sinr-q3-uplink}), since it is the closest position to the side blocks \acp{UE}. %
However, deploying \acp{NCR} in the side blocks is not as good to central block \acp{UE} as the other solutions (\FigRef{FIG:Simulation-results-ncr-sinr-q2-downlink} and \FigRef{FIG:Simulation-results-ncr-sinr-q2-uplink}). %

The solution with two \acp{NCR} in the corners of the central block (scenario 4) seems to have a good balance between improving the \ac{SINR} of side blocks \acp{UE} and the \ac{SINR} of central block \acp{UE}. %
Concerning the \ac{SINR} of side blocks \acp{UE}, this solution is the second best considered solution in both \FigRef{FIG:Simulation-results-ncr-sinr-q3-downlink} and \FigRef{FIG:Simulation-results-ncr-sinr-q3-uplink}. %
Regarding the \ac{SINR} of central block \acp{UE}, this solution performs close to the other solutions with \acp{NCR} in the central block (\FigRef{FIG:Simulation-results-ncr-sinr-q2-downlink} and \FigRef{FIG:Simulation-results-ncr-sinr-q2-uplink}). %
The good balance between central block \acp{UE} and side block \acp{UE} allows this solution to perform as good as the solution with \acp{NCR} in the side blocks. 

Important to highlight that, regarding specifically the central block \acp{UE}, the solution with one \ac{NCR} with two panels pointing toward the side blocks (scenario 3) performs better than the solution with one \ac{NCR} with one panel pointing to the front. %
This is explained by the fact that the solution with two panels covers the central block \acp{UE} positioned in the corners of the central block, while the solution with only one panel pointing towards the front has a weaker coverage in the corners of the central block. %

\begin{table*}[]
	\scriptsize
	\caption{\ac{SINR} increase (in dB) compared to the only macro case at the 10$^{\text{th}}$, 50$^{\text{th}}$ and 90$^{\text{th}}$ percentiles of the SINR CDF.}
	\label{NCR_LOS_TAB:Results-Percentiles2}
	\setlength{\extrarowheight}{1pt}
	\centering
	\begin{tabular}{cc|ccc|ccc|ccc|ccc}
		\cline{3-14}
		&
		&
		\multicolumn{3}{c|}{\textbf{1 NCR}} &
		\multicolumn{3}{c|}{\textbf{1 NCR w/ 2 panels}} &
		\multicolumn{3}{c|}{\textbf{2 NCRs @ center block corners}} &
		\multicolumn{3}{c|}{\textbf{2 NCRs @ side blocks}} \\ \cline{3-14} 
		&
		&
		\multicolumn{3}{c|}{SINR increase (dB)} &
		\multicolumn{3}{c|}{SINR increase (dB)} &
		\multicolumn{3}{c|}{SINR increase (dB)} &
		\multicolumn{3}{c|}{SINR increase (dB)} \\ \hline
		\multicolumn{1}{c|}{\multirow{5}{*}{\rotatebox{90}{Downlink}}} &
		Percentile &
		All UEs &
		central &
		\multicolumn{1}{c|}{side} &
		All UEs &
		central &
		\multicolumn{1}{c|}{side} &
		All UEs &
		central &
		\multicolumn{1}{c|}{side} &
		All UEs &
		central &
		\multicolumn{1}{c|}{side}  \\ \cline{2-14} 
		\multicolumn{1}{c|}{} &
		10$^{\text{th}}$ &
		-2.41 &
		7.43 &
		\multicolumn{1}{c|}{-3.95} &
		-0.29 &
		13.94 &
		\multicolumn{1}{c|}{-2.18} &
		5.21 &
		-1.79 &
		\multicolumn{1}{c|}{12.38} &
		3.04 &
		8.54 &
		\multicolumn{1}{c|}{0.59} \\ \cline{2-14} 
		\multicolumn{1}{c|}{} &
		50$^{\text{th}}$ &
		3.19 &
		12.23 &
		\multicolumn{1}{c|}{-1.79} &
		9.12 &
		14.45 &
		\multicolumn{1}{c|}{3.13} &
		10.91 &
		2.58 &
		\multicolumn{1}{c|}{12.47} &
		11.20 &
		11.17 &
		\multicolumn{1}{c|}{10.65} \\ \cline{2-14} 
		\multicolumn{1}{c|}{} &
		90$^{\text{th}}$ &
		11.81 &
		12.62 &
		\multicolumn{1}{c|}{6.16} &
		13.26 &
		15.94 &
		\multicolumn{1}{c|}{11.59} &
		11.77 &
		5.83 &
		\multicolumn{1}{c|}{13.80} &
		14.48 &
		12.92 &
		\multicolumn{1}{c|}{15.80} \\ \hline
		\multicolumn{1}{c|}{\multirow{5}{*}{\rotatebox{90}{Uplink}}} &
		Percentile &
		All UEs &
		central &
		\multicolumn{1}{c|}{side} &
		All UEs &
		central &
		\multicolumn{1}{c|}{side} &
		All UEs &
		central &
		\multicolumn{1}{c|}{side} &
		All UEs &
		central &
		\multicolumn{1}{c|}{side}  \\ \cline{2-14} 
		\multicolumn{1}{c|}{} &
		10$^{\text{th}}$ &
		0.91 &
		9.81 &
		\multicolumn{1}{c|}{0.05} &
		2.03 &
		13.82 &
		\multicolumn{1}{c|}{0.72} &
		5.57 &
		0.27 &
		\multicolumn{1}{c|}{13.35} &
		5.31 &
		10.28 &
		\multicolumn{1}{c|}{3.49} \\ \cline{2-14} 
		\multicolumn{1}{c|}{} &
		50$^{\text{th}}$ &
		11.04 &
		19.34 &
		\multicolumn{1}{c|}{5.39} &
		15.87 &
		21.41 &
		\multicolumn{1}{c|}{9.93} &
		16.95 &
		8.59 &
		\multicolumn{1}{c|}{18.87} &
		16.07 &
		16.78 &
		\multicolumn{1}{c|}{15.46} \\ \cline{2-14} 
		\multicolumn{1}{c|}{} &
		90$^{\text{th}}$ &
		19.35 &
		20.52 &
		\multicolumn{1}{c|}{11.04} &
		20.25 &
		22.98 &
		\multicolumn{1}{c|}{17.55} &
		19.00 &
		11.12 &
		\multicolumn{1}{c|}{20.65} &
		20.67 &
		18.40 &
		\multicolumn{1}{c|}{23.03} \\ \hline
	\end{tabular}
\end{table*}

\TabRef{NCR_LOS_TAB:Results-Percentiles2} summarizes the \ac{SINR} increase (in dB) due to the deployment of \acp{NCR} compared to the only macro case. %
It presents the \ac{SINR} increase (in dB) at the 10$^{\text{th}}$, 50$^{\text{th}}$ and 90$^{\text{th}}$ percentiles of the \ac{SINR} \ac{CDF} curves of \FigRef{FIG:Simulation-results-ncr-sinr-q2-q3-downlink} and \FigRef{FIG:Simulation-results-ncr-sinr-q2-q3-uplink}. %
Notice that the \ac{UL} direction benefits even more from the deployment of \acp{NCR} than the \ac{DL}. %
This is due to the lower transmit power of the \acp{UE} compared to the transmit power of the \ac{gNB}. %
Moreover, notice that, when considering all the \acp{UE}, the solution with two \acp{NCR} in the corners of the central block is the one with higher gains in the $10^{th}$ percentile for both \ac{DL} and \ac{UL}. %
In other words, this solution benefits the most cell-edge \acp{UE}. %

\section{Conclusions and Future Perspectives}
\label{SEC:Conclusions}

This paper introduced the concept of \ac{NCR}. %
Furthermore, we presented a system level model that allows the performance evaluation of an \ac{NCR}-assisted network. %
Moreover, we evaluated the impact of network deployment on the performance of \ac{NCR}-assisted networks. %

As we showed, network planning considerably affects the performance of \ac{NCR}-assisted networks. %
Moreover, with a proper network planning, the presence of the \acp{NCR} can boost the \ac{SINR} of the \acp{UE} which are in the poor coverage of the \ac{gNB}. %
Furthermore, cell-edge \acp{UE} and \ac{UL} communications are the ones that benefit the most from the presence of \acp{NCR}. %
In the considered scenario, the solution deploying two \acp{NCR} in the corners of the central block presented the best performance, being able to improve the signal quality of \acp{UE} in the central block and in the side blocks. %

Important to highlight that when choosing which solution to deploy, other aspects, not considered in this work, should be considered, e.g. the costs to deploy each considered solution. %




\printbibliography{}
\endrefsection{}
\acbarrier{}

\end{document}